\documentclass[twocolumn]{aastex701}

\begin{document}

\title{
Interplay between Escaping Cosmic Rays and Interstellar Medium: Driving of Galactic Winds and Shaping the Local Proton Spectrum}

\author[0000-0003-3383-2279,sname='Shimoda']{Jiro Shimoda}
\affiliation{Institute for Cosmic Ray Research, The University of Tokyo, 
5-1-5 Kashiwanoha, Kashiwa, Chiba 277-8582, Japan}
\email[show]{jshimoda@icrr.u-tokyo.ac.jp}

%\author[0000-0002-2387-0151,sname='Ohira']{Yutaka Ohira}
%\affiliation{Department of Earth and Planetary Science, The University of Tokyo, 7-3-1 Hongo, Bunkyo-ku, Tokyo 113-0033, Japan}
%\email{y.ohira@eps.s.u-tokyo.ac.jp}

\author[0000-0001-9064-160X,sname='Asano']{Katsuaki Asano}
\affiliation{Institute for Cosmic Ray Research, The University of Tokyo, 
5-1-5 Kashiwanoha, Kashiwa, Chiba 277-8582, Japan}
\email{asanok@icrr.u-tokyo.ac.jp}

\author[0000-0003-4366-6518,sname='Inutsuka']{Shu-ichiro Inutsuka}
\affiliation{Department of Physics, Graduate School of Science, Nagoya University, Furo-cho, Chikusa-ku, Nagoya, Aichi 464-8602, Japan}
\email{inutsuka.shu-ichiro.i2@f.mail.nagoya-u.ac.jp}

%\collaboration{}

%% Use the \collaboration command to identify collaborations. This command
%% takes an optional argument that is either a number or the word "all"
%% which tells the compiler how many of the authors above the command to
%% show. For example "\collaboration[all]{(DELVE Collaboration)}" wil include
%% all the authors above this command.
%%
%% Mark off the abstract in the ``abstract'' environment. 
\begin{abstract}
We study the effects of escaping cosmic rays (CRs)
on the interstellar medium (ISM) around their source
with spherically symmetric CR-hydrodynamical simulations taking
into account the evolution of the CR energy spectrum,
radiative cooling, and thermal conduction.
We show how the escaping CRs accelerate and heat the ISM
depending on the CR diffusion coefficient.
The CR heating effects are potentially responsible for the recent
observations of the unexpected H$\alpha$ and [OIII]$\lambda5007$ lines
in old supernova remnants.
The implied gas outflow rate by CRs can be comparable to the Galactic star
formation rate, compatible with the Galactic wind required for
polluting the halo gas with metals.
Assuming a locally suppressed CR diffusion
and a few nearby CR sources in the Local Bubble, we also propose
alternative interpretations for the Galactic CR proton spectrum
around the Earth measured with CALET, AMS02, and Voyager 1.
\end{abstract}

%% Keywords should appear after the \end{abstract} command. 
%% The AAS Journals now uses Unified Astronomy Thesaurus (UAT) concepts:
%% https://astrothesaurus.org
%% You will be asked to selected these concepts during the submission process
%% but this old "keyword" functionality is maintained in case authors want
%% to include these concepts in their preprints.
%%
%% You can use the \uat command to link your UAT concepts back its source.

%% From the front matter, we move on to the body of the paper.
%% Sections are demarcated by \section and \subsection, respectively.
%% Observe the use of the LaTeX \label
%% command after the \subsection to give a symbolic KEY to the
%% subsection for cross-referencing in a \ref command.
%% You can use LaTeX's \ref and \label commands to keep track of
%% cross-references to sections, equations, tables, and figures.
%% That way, if you change the order of any elements, LaTeX will
%% automatically renumber them.

\section{Introduction}
\label{sec:intro}
The Galactic wind is an important component for the evolution of our Galaxy.
The mass transfer rate of the wind is comparable to the
star formation rate (SFR)
\citep[e.g.,][]{breitschwerdt91,evertt10,recchia16b,shimoda22a},
which implies that the wind controls the star formation in the disk.
Recent observations of external galaxies confirm the
existence of metal-polluted gases around the galaxies
with the concentric radius of up to $\sim100$~kpc
(at the halo or circumgalactic medium, e.g., \cite{tumlinson17}, for review).
The metal-polluted halo strongly indicates the existence of outflows from the disk.
Indeed, the long-term evolution of star formation, metallicity, and stellar dynamics in our galaxy
can be well reproduced by the wind scenario \citep{shimoda24}.
\citet{shimoda_asano24} found that the Galactic wind scenario
possibly explains
the Fermi bubble \citep{su10,su12,ackermann14,sarkar24}
and
eROSITA bubble \citep{predehl20,churazov24,zhang-he-shou24}
as by-products of the interaction between the wind and cosmic rays (CRs).

However, the wind launching mechanism has not been unveiled yet.
The tenuous gas just above the Galactic disk is considered
to be launched by supernova remnants (SNRs, and/or superbubbles), but the gas suffers
radiative cooling, finally, falls back to the disk from a height of $\sim2$ kpc
known as the galactic fountain flow \citep{shapiro76}.
The energy injection from SNRs to the interstellar medium
(ISM) is not so simple,
because almost all kinetic energy is converted to photons
from the shocked plasma 
at a time scale of $\sim50$~kyr \citep{vink12,jimenez19}.
The radius of SNR at $\sim50$~kyr is $\ll100$~pc,
much smaller than the disk thickness. In the case of multiple supernova events
forming a superbubble, the radius is only $\sim80$~pc at $\sim1$~Myr \citep{oku22}.
Moreover, recent observation of NGC~628 (a spiral galaxy similar to the
Milky Way) by {\it James Webb Space Telescope} shows the bubble
size distribution peaked at $\sim30$~pc \citep{watkins23}, significantly
smaller than the theoretical prediction peaked at $\sim100$~pc \citep{nath20}.

In this paper, we focus on the effects of CRs on ISM.
A significant fraction of the kinetic energy of SNRs can also be divided into the CR acceleration.
The accelerated CRs in SNRs eventually escape into the ISM
\citep{ohira10,ohira12}.
In the ISM, the CRs have an energy density of $\sim1$~eV~cm$^{-3}$,
which is comparable with the usual gaseous matter, magnetic field,
and turbulence:
CRs may indirectly support gaseous matter at the height of
$\sim1$~kpc above the midplane of the Galactic disk
\citep{boulares90,ferriere01}.
In the context of the galactic wind, CRs
can prevent the gas from falling and maintain the outflow
toward a far distant height by the CR pressure and heating.

\par
The propagation of escaping CRs is treated as the diffusion process
in the standard picture, but details of interaction between the
escaping CRs and ISM have not been studied well. This can be important
for not only the Galactic evolution but also quantitatively identifying the CR origin.
In this paper, motivated by both issues, we study the interplay
between the escaping CRs and ISM.
\par
Although the diffusion coefficient is the key parameter for CR propagation,
the actual value of the coefficient is a long-standing question.
Averaged arrival time of CRs at energy of $\sim$GeV per nucleon
from the sources to the Earth is estimated from radio isotope ratios such as
the CR $^{10}$Be to $^{9}$Be ratio, implying the arrival time of
$\sim10$--$100$~Myr \citep{ptuskin98}.
On the other hand, observations of CR $^{60}$Fe and $^{59}$Ni
imply much shorter arrival time of $\sim1$~Myr \citep{binns16}.

In the widely accepted picture, CRs distribute from the Galactic disk
to a height of several kpc (i.e., the halo region) \citep[][for reviews]{gabici19}.
The residence time of CRs at the disk is estimated from the primary-to-secondary ratio,
such as the boron-to-carbon ratio.
The diffusion coefficient of $\sim10^{28}~{\rm cm^2~s^{-1}}$ at GeV
that corresponds to several kpc as the scale height of the CR halo,
has been accepted as a fiducial value
\citep[e.g.,][]{strong07,2017PhRvD..95h3007Y}.
Galactic diffuse gamma-ray emission
\citep[e.g.,][]{2012ApJ...750....3A,2025arXiv250218268D}, and
electron+positron CR spectrum
\citep[e.g.][]{2021PhRvD.103h3010E,2022ApJ...926....5A} also suggest
the diffusion coefficient of $\sim10^{28}~{\rm cm^2~s^{-1}}$ at GeV.
This value implies the CR residence time in the Galactic disk $\sim1$--$10$~Myr.
Note that the standard picture basically provides the coefficient at the halo
rather than the disk.
\par
The gas distribution of the Galactic disk may be highly disturbed
and inhomogeneous owing to numerous SNe. While spatially smooth functions
have been assumed for the diffusion coefficient, it can be highly fluctuating. 
Actually, significant suppressions of the diffusion coefficient
($\sim10^{27}~{\rm cm^2~s^{-1}}$ at TeV) have been
reported around pulsar wind nebula with gamma-ray observations
\citep[the so-called TeV-halo, e.g.,][]{hawc17}
and SNR W28 \citep{gabici10}.
Focusing on the above uncertainty in the diffusion
coefficient, we study the effects of escaping CRs on the ISM
by the CR-hydrodynamics system,
taking into account the evolution of the CR distribution,
radiative cooling, and thermal conduction.

\par
This paper is organized as follows. The basic equations of the CR-hydrodynamics
are introduced in section~\ref{sec:cr hydro} and numerical setup
is introduced in section~\ref{sec:set up}.
Although this first-step study makes several simplifications,
we show how the escaping CRs accelerate the ISM and
how the accelerated ISM makes modifications in the CR spectrum
in section~\ref{sec:acc}. The effects of the CR heating are
also studied in section~\ref{sec:heat}.
Then, we discuss
the astrophysical implications of our results, focusing on
the outflow from the Galactic disk in section~\ref{sec:outflow}
and the origin of Galactic CRs in section~\ref{sec:GCR}.
The results are summarized in section~\ref{sec:summary}.

\section{Basic Equations}
\label{sec:cr hydro}
We consider a spherically symmetric system in which the ISM
fluid and CRs coexist. The evolutions of the gas density $\rho_{\rm g}$,
pressure $P_{\rm g}$, and velocity $v_{\rm g}$ are regulated by
the continuity equation, the equation of motion, and the energy
conservation equation as
%
%%%%%%%%%%%%%%%%%%%%%%%%
\begin{eqnarray}
&& \frac{\partial\rho_{\rm g}}{\partial t}
+\frac{1}{r^2}\frac{\partial}{\partial r^2}
\left(r^2\rho_{\rm g}v_{\rm g}\right) = 0,
\label{eq:continuity} \\
&& \rho_{\rm g}\left[
\frac{\partial v_{\rm g}}{\partial t}
+v_{\rm g}\frac{\partial v_{\rm g}}{\partial r}
\right]
=-\frac{\partial}{\partial r}\left(P_{\rm g}+P_{\rm cr}\right),
\label{eq:eom} \\
&& \frac{\partial}{\partial t}
\left(\frac{1}{2}\rho_{\rm g}v_{\rm g}{}^2+\varepsilon_{\rm g}\right)
+\frac{1}{r^2}\frac{\partial}{\partial r^2}
\left[r^2
\left(\frac{1}{2}\rho_{\rm g}v_{\rm g}{}^2
+P_{\rm g}+\varepsilon_{\rm g}\right)v_{\rm g}
\right]
\nonumber \\
&&
=
n_{\rm g}\left(\Gamma_{\rm g}-n_{\rm g}\Lambda\right)
+\frac{1}{r^2}\frac{\partial}{\partial r}
\left(r^2{\cal K}\frac{\partial T_{\rm g}}{\partial r}\right)
\nonumber \\
&&-v_{\rm g}\frac{\partial P_{\rm cr}}{\partial r}
+\int \epsilon
\frac{\partial}{\partial p}
\left[
{\cal N}_{\rm cr}\left(\frac{dp}{dt}\right)_{\rm C}\right]dp
+\big|{\cal V}_{\rm A}
\frac{\partial \varepsilon_{\rm cr}}{\partial r}\big|,
\label{eq:energy gas} 
\end{eqnarray}
%%%%%%%%%%%%%%%%%%%%%%%%
%
respectively, where the gas pressure and
internal energy density $\varepsilon_{\rm g}$ are linked as
%
%%%%%%%%%%%%%%%%%%%%%%%%
\begin{eqnarray}
&& P_{\rm g}=(\gamma_{\rm g}-1)\varepsilon_{\rm g}
=n_{\rm g}kT_{\rm g}=\frac{\rho_g}{\bar{m}}kT_{\rm g},
\label{eq:eos} 
\end{eqnarray}
with the adiabatic index
\begin{eqnarray}
\gamma_{\rm g}=\frac{5}{3}. 
\end{eqnarray}
%%%%%%%%%%%%%%%%%%%%%%%%
Here, we denote the gas temperature and number density
with $T_{\rm g}$ and $n_{\rm g}$, respectively.
The mean molecular mass $\bar{m}$ is assumed to be
the proton mass $m_{\rm p}$ ($\bar{m}$ ranges
$\sim0.6m_{\rm p}$--$1.3m_{\rm p}$, depending on
ionization degree in reality).

\par
In the energy conservation equation (\ref{eq:energy gas}),
we include the radiative heating and cooling terms,
and the thermal conduction term.
The radiative heating rate is set to be
$\Gamma_{\rm g}=2\times10^{26}$ erg s$^{-1}$, following
\citet{koyama02}. The radiative cooling rate, $\Lambda(T_{\rm g})$,
is evaluated under the collisional ionization equilibrium
at $T_{\rm g}\ge10^{4.238}$~K given by \citet{shimoda22a},
and we use the fitting formula at $T_{\rm g}\le10^{4.238}$~K
given by \citet{koyama02}.
The thermal conduction coefficient ${\cal K}$ is given by
\citet{parker53} as ${\cal K}=2.5\times10^3 T_{\rm g}{}^{1/2}$
for $T_{\rm g}\le4.74\times10^4$~K and
${\cal K}=1.25\times10^{-6} T_{\rm g}{}^{5/2}$
for $T_{\rm g}\ge4.74\times10^4$~K.

\par
The terms with the CR pressure $P_{\rm cr}$,
energy density $\varepsilon_{\rm cr}$,
and momentum distribution function ${\cal N}_{\rm cr}(p)$
express the effects of CRs.
The equation of motion (\ref{eq:eom}) includes the force by the CR pressure.
The energy conservation equation (\ref{eq:energy gas}) includes
the mechanical work (adiabatic compression/expansion) by
the CR pressure $-v_{\rm g}\partial_rP_{\rm cr}$, the energy transfer via
the ionization and Coulomb collisions with CRs
$\int \epsilon\partial_p\left[{\cal N}_{\rm cr}(dp/dt)_{\rm C}\right]dp$,
and the heating by the dissipation of Alfv{\`e}n waves
induced by CRs $|{\cal V}_{\rm A}\partial_r \varepsilon_{\rm cr}|$
(introduced later).
\par
We assume that the CR distribution is almost isotropic in the fluid rest frame.
Given the CR momentum distribution function ${\cal N}_{\rm cr}(p)$,
the CR pressure and energy density are defined as
\begin{eqnarray}
P_{\rm cr}=\int \frac{pv}{3} {\cal N}_{\rm cr} dp,~~~
\varepsilon_{\rm cr}=\int \epsilon {\cal N}_{\rm cr} dp,
\end{eqnarray}
where the CR velocity and kinetic energy are functions of momentum $p$ as
\begin{eqnarray}
v(p)&=&\frac{pc^2}{\sqrt{ (m_{\rm p}c^2)^2+(pc)^2 }}, \\
\epsilon(p)&=&\sqrt{ (m_{\rm p}c^2)^2+(pc)^2 }-m_{\rm p}c^2,
\end{eqnarray}
respectively.
The momentum loss of CR protons $(dp/dt)_{\rm C}$ due to pion production,
ionization, and the Coulomb collision is obtained
with the energy loss rate
$(d\epsilon/dt)_{\rm C}$
given by \citet[][see, the equation (5.3.58)]{schlickeiser02} using the relation
$(d\epsilon/dt)_{\rm C}
=(d\epsilon/dp)(dp/dt)_{\rm C}$.
We neglect the pion production process in the CR collisional energy loss.

\par
The heating term $|{\cal V}_{\rm A}\partial_r \varepsilon_{\rm cr}|$
is the consequence of the generation of Alfv{\`e}n waves by CRs
\citep[e.g.,][]{skilling75,achterberg81a,kulsrud05}.
We assume that the generated Alfv{\`e}n waves are immediately
dissipated, which leads to the gas heating
\citep[e.g.,][]{breitschwerdt91, zirakashvili96}.
As we do not solve the evolution of the magnetic field, the Alfv{\`e}n velocity ${\cal V}_{\rm A}=
B_{\rm ism}/\sqrt{4\pi\rho_{\rm g}}$ is calculated with
a fixed value of the field $B_{\rm ism}$,
which corresponds to a system with only a radial component of the magnetic field.
\par
The equation for the CR momentum distribution function
${\cal N}_{\rm cr}(t,r,p)$ is given by
\begin{eqnarray}
&& \frac{\partial {\cal N}_{\rm cr}}{\partial t}
+\frac{1}{r^2}\frac{\partial}{\partial r}
\left( r^2v_{\rm g}{\cal N}_{\rm cr}-
r^2{\cal D}\frac{\partial {\cal N}_{\rm cr}}{\partial r}\right)
\nonumber \\
&&=
\frac{1}{3 r^2}\left(\frac{\partial}{\partial r}
r^2v_{\rm g}\right)
\left( \frac{\partial}{\partial p} p {\cal N}_{\rm cr}\right)
-\frac{\partial}{\partial p}
\left[{\cal N}_{\rm cr} \left(\frac{dp}{dt}\right)_{\rm C}\right] \nonumber \\
&&-\big|{\cal V}_{\rm A}\frac{\partial {\cal N}_{\rm cr}}{\partial r}\big|.
\label{eq:cr trans}
\end{eqnarray}
On the right-hand side, the terms of adiabatic compression/expansion,
collisional momentum loss, and generation of Alfv{\`e}n waves appear.
These energy exchange terms are canceled out in the total energy conservation,
that is, the sum of the contributions from the fluid equation
(\ref{eq:energy gas}) and the CR equation (\ref{eq:cr trans}) integrated by
$\epsilon dp$.
The CR diffusion coefficient ${\cal D}$ is assumed
to be spatially uniform and a function of CR momentum,
${\cal D}={\cal D}(p)$.
As a caveat, the coefficient is
anisotropic in the parallel and perpendicular directions of the magnetic
field in general.
There is likely no particular tendency in the orientation of the magnetic fields.
In our assumption of the small diffusion coefficient due to turbulence, the fields may
be highly entangled. With those uncertainties, we consider only radial diffusion in
our spherically symmetric model. The effects of anisotropic diffusion remain a future topic.

\section{Numerical setup}
\label{sec:set up}
CRs are injected and accelerated at SNR shocks
in the most accepted scenario.
When the radiative cooling of the shock is significant
at say $t\sim50$~kyr \citep[e.g.,][for reviews]{vink12},
we can regard that almost all the supernova energy has been dissipated,
namely the CR injection has been finished. We skip this initial energy
dissipation stage in our simulation.
The CR pressure may be subdominant around the shock front in SNRs.
However, the pressure of escaped CRs in the ISM may affect the ISM motion.
Our initial condition corresponds to the stage when most of CRs have
escaped from the shock front.
We consider an idealized
initial condition: only the CRs are put in a uniform ISM.
From this artificial initial condition, the ISM is accelerated
by the CR pressure and forms a shock wave.
\par
Our initial condition corresponds to the middle-aged SNRs at $t\sim10$~kyr.
The escaping CRs from their source are set to be
%
%%%%%%%%%%%%%%%%%%%%
\begin{eqnarray}
{\cal N}_{\rm cr}(t_{\rm ini},r,p)\propto
\left\{
\begin{array}{cc}
p^{-2.3} & (r\le r_0) \\
0 & (r>r_0)
\end{array}
\right.,
\end{eqnarray}
%%%%%%%%%%%%%%%%%%%%
%
where $r_0=5$ pc and the normalization of $n_{\rm cr}$ is determined by
the total energy of CR as $E_{\rm cr}
=4\pi\int \epsilon n_{\rm cr} r^2drdp=10^{50}$ erg. 
In this paper, we calculate the momentum range of
$10^{-1.5}\le (p/m_{\rm p}c)\le10^{3.5}$,
which is equivalent to
$1~{\rm MeV}\lesssim\epsilon(p)\lesssim3~{\rm TeV}$.
\par
The thermal gas temperature is set to be uniform
as $T_{\rm g}=7000$~K and the gas density is given
by the thermal equilibrium condition of
$n_{\rm g}=\Gamma_{\rm g}/\Lambda(T_{\rm g})\simeq0.2$~cm$^{-3}$
($P_{\rm g}\simeq0.1~{\rm eV~cm^{-3}}$).
The sound speed is $\simeq10~{\rm km~s^{-1}}
(T_{\rm g}/7000~{\rm K})^{1/2}$.
As will be shown, the accelerated flow becomes supersonic.
The number density is consistent with
the space-averaged number density around the solar system
\citep[see,][for reviews]{ferriere01}.
We consider a spatial radius range as
$0~{\rm pc}\le r\le300$~pc, in which the total thermal energy
is initially $E_{\rm g}\simeq10^{51}$ erg and the
total mass is $M_{\rm g}\simeq5.7\times10^5~M_\odot$. 
The initial gas velocity is $v_{\rm g}(t_{\rm ini},r)=0$ everywhere.
The gas heating by the CR-induced Alfv{\`e}n wave is considered
under a fixed magnetic field strength of $B_{\rm ism}=1~{\rm \mu G}$,
for simplicity.
This parameter affects only the gas heating rate.
The field strength of $\sim1~{\rm \mu G}$ is consistent with
that estimated from the full-sky rotation measure analysis as a typical value of
the disk \citep{unger24}. However, as caveats, it may spatially vary in multiphase
ISM, e.g., in molecular clouds, or due to magnetic field amplification by CR streaming
\citep[e.g.,][]{bell04}. Even in the case of HII regions, such as S235 complex,
the estimated strength can be up to $\sim50~{\rm \mu G}$ at clumps \citep{devaraj21}.
Such local enhancements of the field may significantly change the estimates of
CR heating rate.
\par
The CR diffusion coefficient is one of the most uncertain
parameters. In this paper, for simplicity, we set spatially
uniform coefficients as
%
%%%%%%%%%%%%%%%%%%%
\begin{eqnarray}
{\cal D}(p)={\cal D}_0\left(\frac{p}{m_{\rm p}c}\right)^{\delta}.
\label{eq:diff coeff}
\end{eqnarray}
%%%%%%%%%%%%%%%%%%%
%
%where we adopt $a=1/3$ (Kolmogorov turbulence) in this paper.
Diffusion models in previous studies succeed in fitting the CR spectra,
including the boron-to-carbon ratio, with various values of the index $\delta$
from the Kolmogorov value $1/3$ \citep[e.g.,][]{2019PhLB..789..292W} to a larger value,
such as $0.7$ \citep[e.g.,][]{2022PhRvL.129y1103A}. As the index $\delta$ in a local ISM
with a suppressed diffusion coefficient is uncertain, we adopt the Kolmogorov value $\delta=1/3$.
The Kolmogorov-like turbulence is likely for ionized medium as implied by
Voyager 1 and 2 observations \citep[e.g.,][]{burlaga15a,burlaga15b}. The observation of NGC~628
implies ionized bubbles occupy a large volume fraction of the galactic disk \citep{watkins23}.
We parameterize our calculation by the representative
value of ${\cal D}_0$ in the range of
$10^{26}$--$10^{28}$ cm$^2$ s$^{-1}$, considering the locally suppressed coefficient.
In the most noteworthy case of ${\cal D}_0=10^{26}~{\rm cm^2~s^{-1}}$,
we also show the numerical result
for $10^{-1.5}\le (p/m_{\rm p}c)\le10^{5.5}$,
which is equivalent to
$1~{\rm MeV}\lesssim\epsilon(p)\lesssim300~{\rm TeV}$.

\section{Temporal Evolution of ISM Driven by Escaping Cosmic Rays}
\label{sec:temp}
Here, we describe the temporal evolution of ISM and CRs.
Firstly, the acceleration of the fluid by the CR pressure
is discussed. Then, the CR heating effect and its
observational test are discussed.\footnote{
The simulation movies are available at
\url{https://youtube.com/playlist?list=PLgnUM4yGp9oLt03moYzVb8DJzfSDrw9Ft&si=BxyjXfEBIw3_Q-r5} }

\subsection{Acceleration of the fluid and deformation
of the cosmic ray spectrum}
\label{sec:acc}
\begin{figure}
\plotone{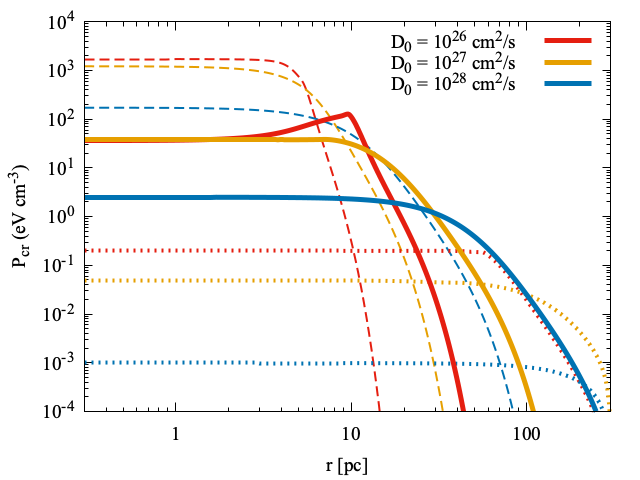}
\caption{
The CR pressure profiles for the cases of
${\cal D}_0=10^{26}~{\rm cm^2~s^{-1}}$ (red),
$10^{27}~{\rm cm^2~s^{-1}}$ (orange), and
$10^{28}~{\rm cm^2~s^{-1}}$ (blue).
The dashed, solid, and dotted lines are the
profiles at $t=0.5$~kyr, $10$~kyr, and $1$~Myr,
respectively.}
\label{fig:r_Pcr}
\end{figure}

\begin{figure}
\plotone{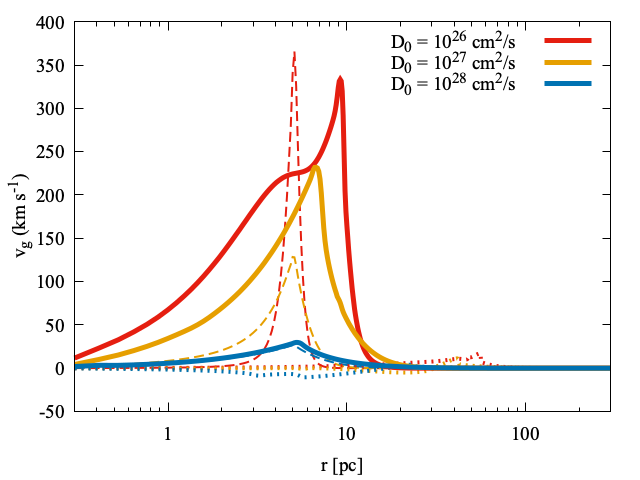}
\caption{
The fluid velocity profiles for the cases of
${\cal D}_0=10^{26}~{\rm cm^2~s^{-1}}$ (red),
$10^{27}~{\rm cm^2~s^{-1}}$ (orange), and
$10^{28}~{\rm cm^2~s^{-1}}$ (blue).
The dashed, solid, and dotted lines are the
profiles at $t=0.5$~kyr, $10$~kyr, and $1$~Myr,
respectively.
}
\label{fig:r_vg}
\end{figure}

\begin{figure}
\plotone{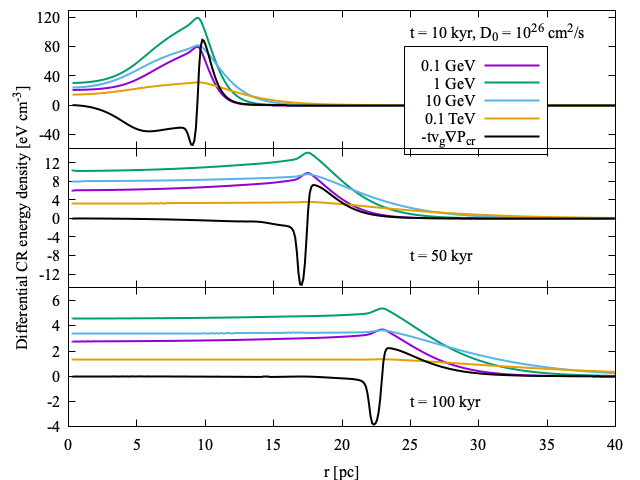}
\caption{
Temporal evolutions of the differential CR energy density,
$\epsilon p{\cal N}_{\rm cr}$, for the case of
${\cal D}_0=10^{26}~{\rm cm^2~s^{-1}}$.
The energy transfer by the mechanical work by CRs,
$-tv_{\rm g}\partial_rP_{\rm cr}$, is
also shown by the black curves.
The positive $-tv_{\rm g}\partial_rP_{\rm cr}$
indicates the energy transfer from CRs to the fluid, while the negative value indicates the energy gain of CRs via the compression
of the fluid.}
\label{fig:cr work}
\end{figure}

Figure~\ref{fig:r_Pcr} and \ref{fig:r_vg} show
the radial profiles of $P_{\rm cr}$ and $v_{\rm g}$, respectively.
The early stage evolution is highly affected by our artificial
initial condition. Although the behavior in this stage is not of
our interest, the figures clearly show the dependence of the diffusion
coefficient. A smaller ${\cal D}_0$ leads to a steeper 
profile of the CR pressure, which accelerates the ISM fluid to
a higher velocity.  As shown in the figure, the gas velocity is
supersonic (the sound speed is 10 km $\mbox{s}^{-1}$).
The CR pressure bump seen for
${\cal D}_0=10^{26}~{\rm cm^2~s^{-1}}$ at $t=10$~kyr is 
due to the back-reaction from the accelerated ISM gas.
The compressed CRs are heated  in this region. This heating effect
is prominent in the very early stage, as shown in 
Figure \ref{fig:cr work}. Here, the differential CR
energy density, $\epsilon p{\cal N}_{\rm cr}$, and the transferred
energy due to the mechanical work $-tv_{\rm g}\partial_rP_{\rm cr}$
are plotted for the case of ${\cal D}_0=10^{26}~{\rm cm^2~s^{-1}}$.
The CR energy is transferred to the fluid in the region with
$-tv_{\rm g}\partial_rP_{\rm cr}>0$. While in the region with
$-tv_{\rm g}\partial_rP_{\rm cr}<0$, the fluid adiabatically
compresses the CRs, transporting lower energy CRs to a higher energy part.

\par
%%%%%%%%%%%%%%%%%%%%
%%%%%%%%%%%%%%%%%%%%
%
%
%%%%%%%%%%%%%%%%%%%%
%%%%%%%%%%%%%%%%%%%%
%
\par
%
%%%%%%%%%%%%%%%%%%%%
%%%%%%%%%%%%%%%%%%%%
%
%
%%%%%%%%%%%%%%%%%%%%
\begin{figure}
\plotone{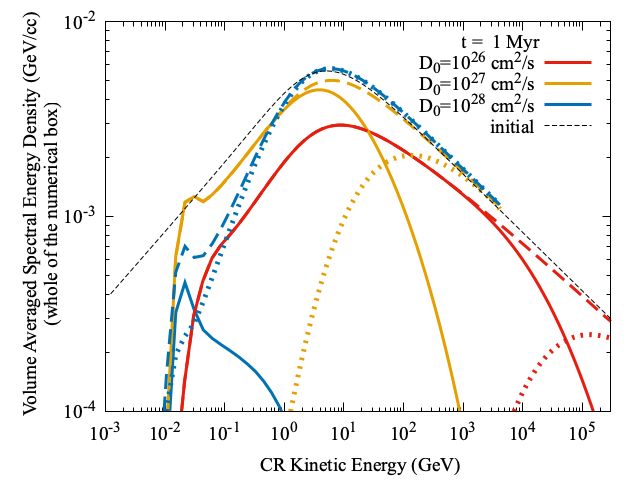}
\plotone{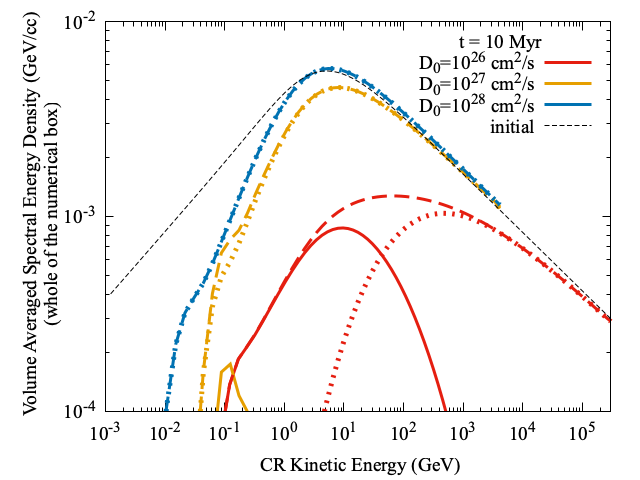}
\caption{The solid lines are the volume averaged
CR energy spectrum in
the whole simulation box
($r\leq300$~pc) at $t=1$~Myr (upper) and $10$~Myr (lower) for the cases of
${\cal D}_0=10^{26}~{\rm cm^2~s^{-1}}$ (red),
$10^{27}~{\rm cm^2~s^{-1}}$ (orange), and
$10^{28}~{\rm cm^2~s^{-1}}$ (blue).
The dotted lines are spectra of the CRs escaped from the simulation box.
The dashed thick lines are the sum of the escaped and remaining components.
The thin dashed line (black) shows the initial spectrum common to all the cases.}
\label{fig:hardening}
\end{figure}
%%%%%%%%%%%%%%%%%%%%
%
\par
However, in the late phase, which is our main subject of interest,
the mechanical energy loss for CRs is dominant rather than the energy gain.
Figure~\ref{fig:hardening} shows the volume averaged
CR energy spectra at $t=1$~Myr and 10 Myr.
The solid and dotted lines are spectra of the remaining CRs
in the simulation box and the escaped CRs, respectively. 
The escaped CR spectrum is calculated by time integration
of the spatial flux,
$\partial_r[r^2v_{\rm g}f-{\cal D}\partial_rf]/r^2$.
For ${\cal D}_0=10^{28} {\rm cm^{2}~s^{-1}}$ and $10^{27} {\rm cm^{2}~s^{-1}}$,
a significant fraction of higher energy CRs escape from the simulation box
at $t=1$ Myr. This is consistent with the escape timescale
$(300 {\rm pc})^2/(4{\cal D})\sim 1$ Myr at 1 GeV for ${\cal D}_0=10^{28}
{\rm cm^{2}~s^{-1}}$. The low energy deficit (below 0.1 GeV) is due to
the collisional energy loss.
For the suppressed diffusion coefficient (${\cal D}_0=10^{26} {\rm cm^{2}~s^{-1}}$), 
the slow diffusion still confines most of CRs in the calculation box. The steep 
profile of the CR pressure causes a significant energy transfer
to the ISM fluid. This  energy loss is reflected as the difference between
the initial (thin black dashed  line) and total (red dashed line) spectra.
\par
At $t=10$ Myr, all CRs have escaped from the simulation box
for ${\cal D}_0=10^{28} {\rm cm^{2}~s^{-1}}$ and $10^{27} {\rm cm^{2}~s^{-1}}$.
The final spectra of the escaped CRs are not significantly modified.
This is consistent with the standard assumption for the global-scale
CR injection. Even in the case of ${\cal D}_0=10^{26} {\rm cm^{2}~s^{-1}}$,
only 1--10 GeV CRs remain in the simulation box.
The low flux of the total CR spectrum 
(red dashed line) suggests further energy transfer to the fluid during $t=1$--10 Myr.
As lower-energy CRs, whose density profile is steeper, tend to work more
efficiently on the fluid before escape, the higher energy loss for
the lower-energy CRs results in a harder spectrum than the initial spectrum.

\par
The simple estimate of the diffusion timescale
$(300 {\rm pc})^2/(4{\cal D}_0)$ leads to $\sim 100$ Myr
for ${\cal D}_0=10^{26}~{\rm cm^2~s^{-1}}$. However, as shown in
Figure~\ref{fig:hardening}, most of CRs lose their energy before escape.
The effective residence time is much shorter than the escape timescale.
This reduction of the residence time may lead to the universal residence time
of 1-10 Myr implied from the boron-to-carbon ratio, despite
the inhomogeneity of the diffusion coefficient in our galactic disk.
We discuss the observed CR proton spectrum around the Earth
using the results of ${\cal D}_{0}=10^{26}~{\rm cm^2~s^{-1}}$ later
in section~\ref{sec:GCR}.
\subsection{Cosmic ray heating}
\label{sec:heat}
\begin{figure}
\plotone{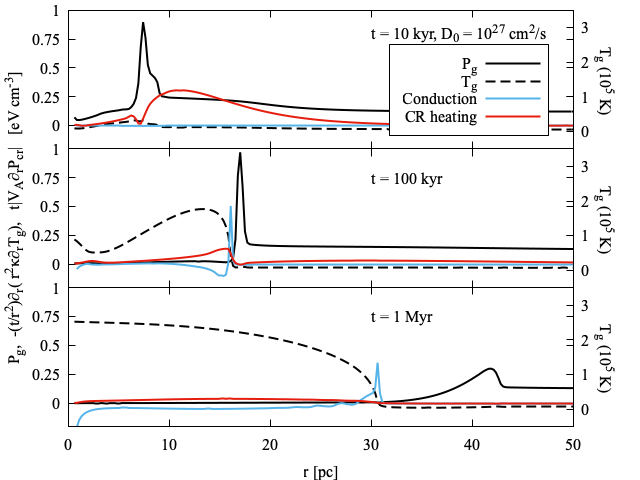}
\caption{Temporal evolutions of
the pressure $P_{\rm g}$ (the solid black curve) and
temperature $T_{\rm g}$ (the dashed black curve)
for ${\cal D}_0=10^{27}~{\rm cm^2~s^{-1}}$.
We also exhibit the CR heating term
$t|{\cal V}_{\rm A}\partial_r\varepsilon_{\rm cr}|$ (red)
and the thermal conduction term
$-(t/r^2)\partial_r\left(r^2{\cal K}\partial_rT_{\rm g}\right)$
(lightblue).}
\label{fig:conduction}
\end{figure}

The gas heating term via the 
dissipation of the Alfv{\`e}n waves induced by CRs,
$|{\cal V}_{\rm A}\partial_r\varepsilon_{\rm cr}|
\sim{\cal V}_{\rm A}\varepsilon_{\rm cr}/\sqrt{4{\cal D}_0 t}$,
is effective at the diffusion front.
The steeper pressure profile for smaller values of ${\cal D}_0$ enhances
the fraction of the CR energy loss with this dissipative mechanism.
The total energy transferred to the thermal plasma is
evaluated as
%
%%%%%%%%%%%%%%%%%%%%
\begin{eqnarray}
\Delta Q_{\rm cr,w}
&& \sim t{\cal V}_{\rm A}\frac{E_{\rm cr}}{\sqrt{4\pi{\cal D}_0t}}
\nonumber \\
&& \sim 0.24E_{\rm cr}
\left(\frac{n_{\rm g}  }{0.2~{\rm cm^{-3}}}\right)^{-1/2}
\left(\frac{B_{\rm ism}}{1~{\rm \mu G}}\right)
\nonumber \\
&&\times
\left(\frac{t}{10~{\rm Myr}}\right)^{1/2}
\left(\frac{{\cal D}_0 }{10^{26}~{\rm cm^2 s^{-1}}}\right)^{-1/2},
\end{eqnarray}
%%%%%%%%%%%%%%%%%%%%
%
which is not so significant for the thermal plasma
with the initial total energy of $E_{\rm g}\simeq
10^{51}~{\rm erg}\simeq10~E_{\rm cr}$.
Even with a higher $B_{\rm ism}$ and smaller ${\cal D}_0$,
the maximum $\Delta Q_{\rm cr,w}$ is limitted to $E_{\rm cr}\sim 0.1 E_{\rm g}$ or less.
However, the heating rate is significant
at the cavity formed by the expansion.
As shown in Figure~\ref{fig:conduction},
for the case of ${\cal D}_0=10^{27}~{\rm cm^2~s^{-1}}$, the CR heating
is effective outside the expanding shell initially ($t\lesssim10$~kyr).
Even with the conservative magnetic field of $B_{\rm ism}=1\mu$G,
the temperature increases to
$\sim10^5$~K by taking a time of $\sim100$~kyr, while the expansion
flow is supersonic as shown in the Figure \ref{fig:r_vg}.
Thus, the high-temperature gas has been left in the cavity.
At $t\gtrsim100$~kyr, the thermal conduction rate becomes comparable with
the heating rate. The conduction smoothes out the temperature structure.
%
%%%%%%%%%%%%%%%%%%%%
%%%%%%%%%%%%%%%%%%%%
%
\par
%
%%%%%%%%%%%%%%%%%%%%
\begin{figure}
\plotone{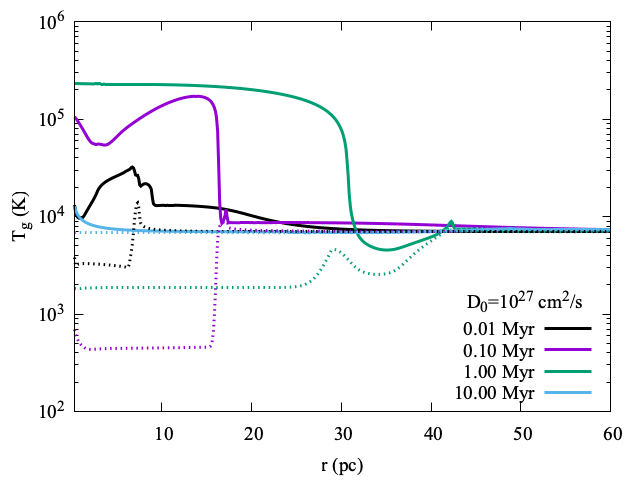}
\caption{
The comparison of the temperature profiles
in the cases with (solid) and without (dots) the CR heating  
for ${\cal D}_0=10^{27}~{\rm cm^2~s^{-1}}$.
}
\label{fig:heating}
\end{figure}
%%%%%%%%%%%%%%%%%%%%
%
Figure~\ref{fig:heating} shows a comparison of the temperature
evolution between the calculations with and without the CR heating
effect. A large difference in the temperature inside the expanding
region (cavity) is shown in the figure.
Note that the heating due to
the particle-particle collisions, $(d\epsilon/dt)_{\rm C}$,
is not efficient inside the cavity.
\par
%
%%%%%%%%%%%%%%%%%%%%
\begin{figure}
\plotone{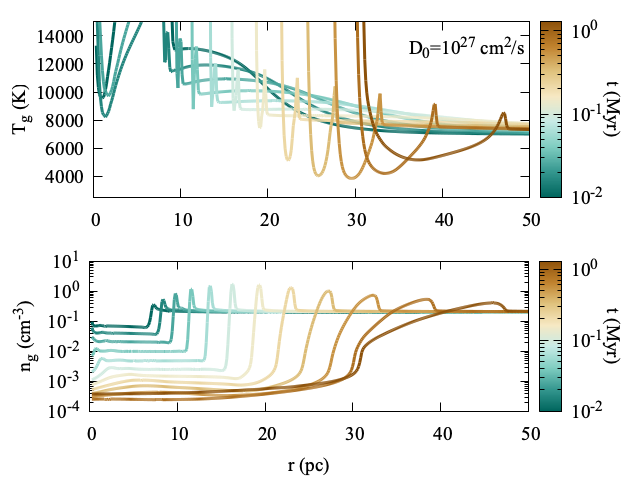}
\caption{The evolutions of the gas temperature (top) and
number density (bottom) profiles from $t=10$ kyr to
1 Myr for ${\cal D}_0=10^{27}~{\rm cm^2~s^{-1}}$.}
\label{fig:bump}
\end{figure}
%%%%%%%%%%%%%%%%%%%%
%
The effects of the CR heating with the supersonic expansion
also results in a characteristic temperature structure
around the edge of the cavity. Figure \ref{fig:bump} is
the close-up views of the temperature and number density profiles
for ${\cal D}_0=10^{27}~{\rm cm^2~s^{-1}}$.
At $t=1$ Myr, the energy transferred from CRs is
$\Delta Q_{\rm cr,w}\sim0.08~E_{\rm cr}$.
The temperature shows a large jump at
$(t,r)\sim(1~{\rm Myr}, 30~{\rm pc})$ corresponding to
the edge of the cavity and a small jump at
$(t,r)\sim(1~{\rm Myr}, 45~{\rm pc})$
reflecting the shock.
From the shock to the edge, the temperature decreases gradually.
We refer to the low-temperature region as the tail,
while the shock heated region as the bump.
\par
The bump and tail are possibly observational
counterparts of the effects of CR heating by
bright atomic lines such as
H$\alpha$ at $6000$~K$\lesssim T_{\rm g}\lesssim10^{4}$~K
and [OIII]$\lambda5007$
at $10^4$~K$\lesssim T_{\rm g}\lesssim10^{5}$~K
\citep[e.g.,][]{osterbrock06}.
If the case, the [OIII]$\lambda5007$ is bright outside (bump),
while the H$\alpha$ is bright inside (tail).
Interestingly, \citet{fesen24} reported such
[OIII]$\lambda5007$ and H$\alpha$ at old SNRs:
[OIII]$\lambda5007$ filaments are bright at the outer side
than H$\alpha$ filaments.
The presence of this outside high-ionized region is not trivial.
The heating by CRs plays an important role in this feature.
The projection effect can reproduce the filamentary surface brightness
profiles.
Note that such a bump-tail structure almost vanishes
when the diffusion coefficient is large, such as
${\cal D}_0=10^{28}~{\rm cm^2~s^{-1}}$.
\par
The observed features for [OIII] and H$\alpha$ could potentially
be explained by other mechanisms
such as combinations of radiative shock precursors
\citep{1981PASJ...33....1I,1994ApJ...420..268C},
or the projection effects of shock rippling. 
However, the simple plane-parallel shock model predicts that
the ionization degree increases from far upstream to downstream
\citep[e.g.,][]{sutherland17}. This trend is contrary to the
observed feature of a clear separation
\citep[figure 7 of][is one of the representatives]{fesen24}.
Shock waves are unable to form an
extensive photoionization precursor responsible for the more
distant [OIII] emission.
If the observed separation of [OIII] and H$\alpha$ is intrinsic
at the sources, the presence of [OIII] emission region outside H$\alpha$
implies a local heating of gaseous matter.
Our results suggest that the escaping CR heating forms such
a complex structure.
We will study the observational counterparts of
the CR effects along such lines in the future.
\par
%
%%%%%%%%%%%%%%%%%%%%
\begin{figure}
\plotone{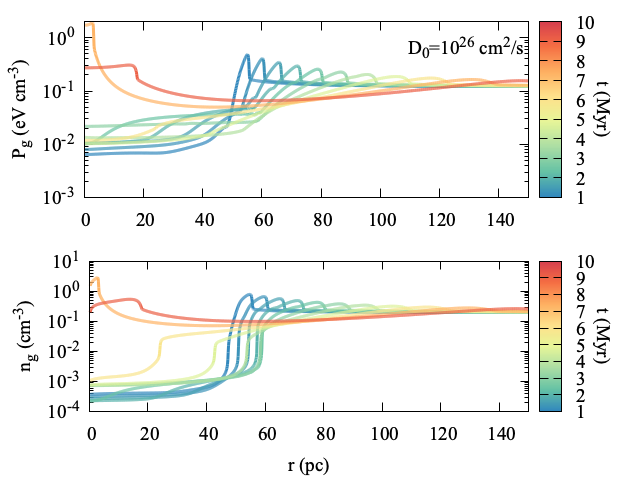}
\caption{The evolutions of the gas pressure (top) and number density (bottom)
profiles from $t=1$ Myr to 10 Myr for ${\cal D}_0=10^{26}~{\rm cm^2~s^{-1}}$.
At $t\gtrsim6$~Myr, the expanded fluid blows back due to
the decreasing pressure inside the cavity.
}
\label{fig:blow_back}
\end{figure}
%%%%%%%%%%%%%%%%%%%%
%
%
%%%%%%%%%%%%%%%%%%%%
\begin{figure}
\plotone{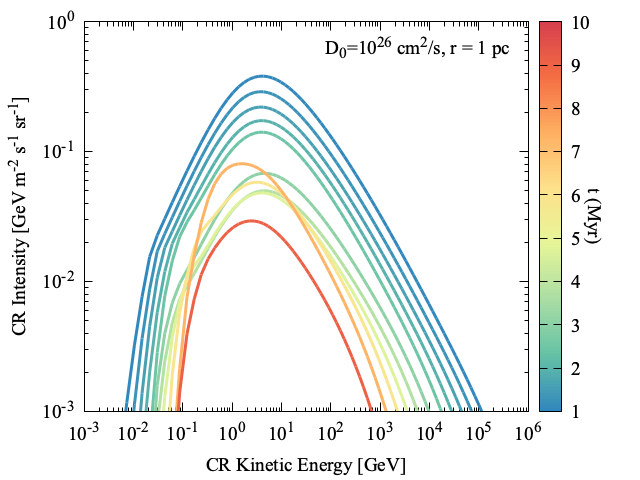}
\caption{The temporal evolution of
the CR intensity spectrum
for ${\cal D}_0=10^{26}~{\rm cm^2~s^{-1}}$
at $r=1$~pc.
At $t\sim6$~Myr, the CR intensity temporarily increases due to
the blow-back of fluid.}
\label{fig:local_cr}
\end{figure}
%%%%%%%%%%%%%%%%%%%%
%
When the pressure in the cavity is sufficiently small,
the ambient fluid begins to blow back and finally fills
the cavity (Figure~\ref{fig:blow_back}). In the case of
a small diffusion coefficient such as
${\cal D}_0\sim10^{26}~{\rm cm^2~s^{-1}}$, the low-energy
CRs are transported by the backflow.
Figure~\ref{fig:local_cr} shows the temporal evolution
of the CR intensity at $r=1$~ pc for
${\cal D}_0=10^{26}~{\rm cm^2~s^{-1}}$.
Initially, the CR intensity decreases as the inner cavity evolves.
At $t\sim6$~Myr, the local CR intensity temporarily increases 
due to the backflow of the fluid.

\section{The Astrophysical Implications}
\label{sec:implication}
We discuss the implications of our results concerning
the outflow from the Galactic disk in section~\ref{sec:outflow}
and the observed Galactic CR spectrum around the Earth
in section~\ref{sec:GCR}.

\subsection{Outflow from the Galactic disk}
\label{sec:outflow}
%
%%%%%%%%%%%%%%%%%%%%
\begin{figure}
\plotone{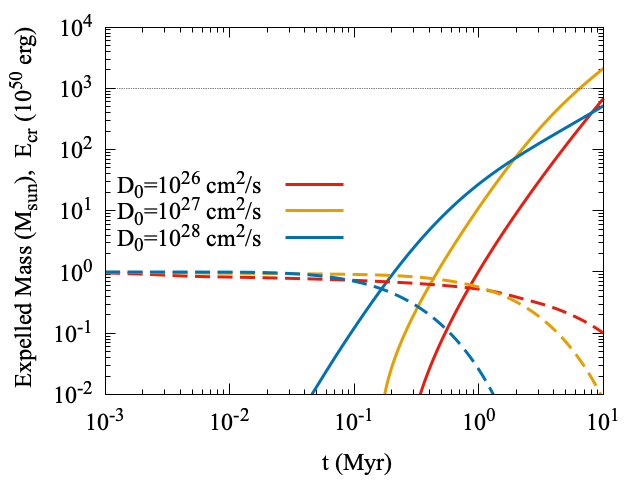}
\caption{The total mass expelled by CRs as a function of time.
The horizontal dots indicate the expelled mass
of $10^3~M_\odot$ as a guide. The dashed lines are
the total CR energy in the calculation box.
}
\label{fig:mass_loss}
\end{figure}
%%%%%%%%%%%%%%%%%%%%
%
In Figure~\ref{fig:mass_loss}, we show how much mass
is expelled from the $200$~pc sphere
(the numerical boundary is at $300$~pc).
We also plot the total CR energy in the calculation box. 
For ${\cal D}_0=10^{27}~{\rm cm^2~s^{-1}}$ and
$10^{28}~{\rm cm^2~s^{-1}}$, the final 
energy decrease is mainly due to escape of CRs, while
the energy transfer to the fluid 
is the reason for $10^{26}~{\rm cm^2~s^{-1}}$.
The coefficient of ${\cal D}_0=10^{27}~{\rm cm^2~s^{-1}}$
results in the most significant mass loss. In the standard case of
${\cal D}_0=10^{28}~{\rm cm^2~s^{-1}}$, the available
CR energy is small due to the escape of CRs, leading to a lower mass loss.
On the other hand, for
${\cal D}_0=10^{26}~{\rm cm^2~s^{-1}}$, 
the slow diffusion may slightly suppress the mass ejection rate.
However, the differences in the expelled mass are only a factor of 2 around $\sim10^3~M_\odot$,
so that the mass loss
does not sensitively depend on the diffusion coefficient.
\par
The expelled mass of $M_{\rm ex}\sim10^3~M_\odot$ per one CR source
has a significant impact on the long-term evolution of
the Galactic disk. Supposing that supernovae are typical CR sources, as usual,
the supernova rate of
$\dot{N}_{\rm sn}\sim0.01~{\rm yr^{-1}}$
results in a disk mass loss rate due to the outflow of
$\dot{N}_{\rm sn}M_{\rm ex}\sim10~M_\odot~{\rm yr^{-1}}$.
This is comparable to the SFR in our Galaxy
\citep{haywood16}. Indeed, \citet{shimoda24} shows that
the star formation history in the Milky Way over cosmic time can be
well reproduced by assuming such outflows.
\par
The results imply that CRs transfer
gas and metals from the disk to the halo. This does not contradict
the appearance of disk galaxy such as NGC~628 that is full of
small bubbles \citep[$\sim30$~pc][]{watkins23}.
Once the ISM goes to the halo, its tenuous components can be the Galactic
wind as shown by the previous studies \citep[e.g.,][]{breitschwerdt91}.
The Galactic wind should be responsible for the metal pollution of the halo
at a height of $\sim100$~kpc as seen in external galaxies \citep{tumlinson13,tumlinson17,shimoda22a}.
\par
To study the Galactic outflow in detail, we should extend our model
by including the effects of the stratification of the disk
gas (effects of gravity), supernova blast waves, magnetic field,
and so on. In particular, \citet{breitschwerdt99} points out
the importance of the disk-halo interface for understanding the Galactic wind
\citep[see, also][for one of the latest simulations]{habegger25,armillotta25}.
This would also test the recently suggested scenario of the origin
of Fermi and eROSITA bubbles by \citet{shimoda_asano24}.
Toward establishing a comprehensive picture of the Galactic wind, we will study the outflows
along such lines in the future.

\subsection{Local cosmic ray spectrum}
\label{sec:GCR}
In the standard scenario, the CR spectrum observed around
the Earth is a superposition of spectra from distant sources.
For the standard value of the diffusion coefficient
${\cal D}_0\sim10^{28}~{\rm cm^2~s^{-1}}$, the typical distance of
sources of $\sim1$~GeV protons
is $\sim\sqrt{4{\cal D}_0 \tau_{\rm res}}\sim1~{\rm kpc}$
$({\cal D}_0/10^{28}~{\rm cm^2~s^{-1}})^{1/2}$
$(\tau_{\rm res}/10~{\rm Myr})^{1/2}$, where
$\tau_{\rm res}\sim1$--$10$~ Myr is the representative residence time
of CRs in the Galaxy inferred from the composition of CR isotopes.
Our result with ${\cal D}_0=10^{28}~{\rm cm^2~s^{-1}}$ shows that
the spectrum of the escaped CRs is almost not affected by the interaction
with ISM gases. This justifies the standard scenario.
\par
However, as the solar system is within the Local Bubble
\citep{1987ARA&A..25..303C,zucker22}, which is a remnant of multiple
supernovae, a locally suppressed
diffusion coefficient is an attractive possibility to
consider the local CR spectrum.
If the coefficient is ${\cal D}_0=10^{26}~{\rm cm^2~s^{-1}}$
as implied from TeV Halos \citep[e.g.,][]{hawc17,giacinti20,amato24},
the typical source distance becomes $\sqrt{4{\cal D}_0 \tau_{\rm res}}
\sim100~{\rm pc} ({\cal D}_0/10^{26}~{\rm cm^2~s^{-1}})^{1/2}
(\tau_{\rm res}/10~{\rm Myr})^{1/2}$ as mentioned in section~\ref{sec:acc}.
The distance is comparable to the Local Bubble.
The solar system is considered to have crossed the edge of the Local Bubble $\sim6$~Myr ago by its proper motion and is now located around the center \citep{zucker22}.
The idea of the suppression of the diffusion coefficient
around CR sources is given by \citet{cowsik73,cowsik75} and
has been under debate \citep[][as one of the latest]{schroer25}.

\par
Such nearby source scenarios have also been studied in the literature.
Recent observations of short-lived radioactive nuclei in CRs suggest
that the bulk of low-energy CR
comes from the modest number of supernovae occurring several Myr ago,
which are currently part of the Local Bubble
\citep[e.g.,][]{erlykin12,boschini21,shi25}.
The Combination of
$^{59}$Ni with a half-time of $76$~kyr and
$^{60}$Fe with a half-time of $2.6$~Myr is
one of the representatives \citep{binns16},
leading to a mean time between nuclear synthesis
and their arrival as
$0.1~{\rm Myr}\lesssim t\lesssim$ several Myr.
The nearby supernova activities
are also implied by the composition of ocean crusts:
$^{60}$Fe implies active phases at
$\sim1$--$3$~Myr ago and $\sim6$--$7$~Myr ago \citep{wallner21}.
The other analysis of $^{10}$Be implies $\sim10$~Myr ago
\citep{koll25}.
\par
%
%%%%%%%%%%%%%%%%%%%%
\begin{figure}
\plotone{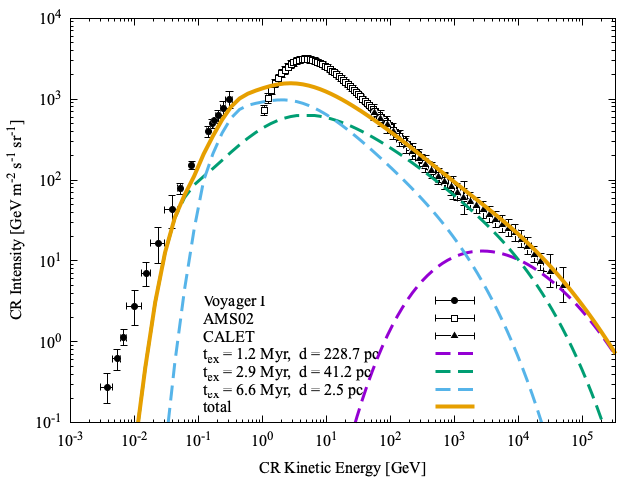}
\caption{The spectral intensity of the observed proton CRs
around the Earth AMS-02 \citep{aguilar15} and CALET \citep{adriani22}
and beyond the termination shock of the solar wind
\cite[Voyager 1][]{cummings16}.
The dashed lines are the intensities of CRs from
the source with age of $t_{\rm age}$ and distance of $d$
from the Earth, assuming ${\cal D}_0=10^{26}~{\rm cm^2~s^{-1}}$.
The orange solid line represents the sum of
the contributions from three distinct sources as
$(t_{\rm age},d)=(1.2~{\rm Myr},~ 228.7~{\rm pc})$ (purple),
$(2.9~{\rm Myr},~  41.2~{\rm pc})$ (green), and
$(6.6~{\rm Myr},~   2.5~{\rm pc})$ (cyan).}
\label{fig:GCR}
\end{figure}
%%%%%%%%%%%%%%%%%%%%
%
Here, we demonstrate reproducing the CR proton spectrum observed at
the Earth with this nearby source scenario. Our purpose is to propose
new possible scenarios rather than a detailed fitting by adjusting
multiple parameters, which cannot be determined uniquely.
As an example, we consider three distinct sources:
each source ejects CRs $1.2$~ Myr ago, $2.9$~ Myr ago, and $6.6$~ Myr ago,
respectively. Note that the enhancement of $^{60}$Fe in the ocean crusts
at $\sim1$--$3$~Myr looks more drastic than one at $\sim6$--$7$~Myr.
Then, we parameterize the distance of the source center ($r=0$~pc)
from the current position of the solar system, $d=r$.
\par
Figure~\ref{fig:GCR} is a dominantly contributing case
that roughly reproduces the measured proton spectrum,
assuming ${\cal D}_0=10^{26}~{\rm cm^2~s^{-1}}$. 
The small diffusion coefficient in the Local Bubble may block
the penetration of CRs from distant sources, and CRs from the local
($\sim 100$ pc) sources may dominate in the observed spectrum as
demonstrated here.
In this case, the sources are assumed to be at
$(t_{\rm age},d)=(1.2~{\rm Myr},~ 228.7~{\rm pc})$,
$(t_{\rm age},d)=(2.9~{\rm Myr},~  41.2~{\rm pc})$, and
$(t_{\rm age},d)=(6.6~{\rm Myr},~   2.5~{\rm pc})$, respectively.
Interestingly, the variety of the individual spectra due to
the interaction with the ISM can produce a total spectrum consistent
with the observed hardening at $\sim 600$ GeV and softening at $\sim 10$ TeV.
\par
The gas column density measured along a CR particle trajectory,
called the ``grammage'', is evaluated from the CR boron-to-carbon ratio
as $\Lambda_{\rm gr}\sim10~{\rm g~cm^{-2}}$,
where the boron is created via nuclear spallations.
In our case, the average grammage is estimated as
$\Lambda_{\rm gr}\sim\rho_{\rm g}ct_{\rm age}
\sim2~{\rm g~cm^{-2}}$
$(n_{\rm g,ini}/0.2~{\rm cm^{-3}})$
$(t_{\rm age}/6.6~{\rm Myr})$.
If we take into account the backflow effects, the low-energy
CRs are convected from the compressed
region (Figures~\ref{fig:blow_back} and \ref{fig:local_cr}),
we may obtain an enhanced one as
$\Lambda_{\rm gr}
\sim10~{\rm g~cm^{-2}}$
$(n_{\rm g}/1~{\rm cm^{-3}})$
$(t_{\rm age}/6.6~{\rm Myr})$.
The predictions in the model in Figure~\ref{fig:GCR}
are the energy-dependent CR ``age'':
$t_{\rm age}=1.2$~Myr at $\epsilon\gtrsim10$~TeV,
$t_{\rm age}=2.9$~Myr at $\epsilon\lesssim100$~MeV
and $100$~GeV$\lesssim\epsilon\lesssim10$~TeV, and
$t_{\rm age}=6.6$~Myr at
$100$~MeV$\lesssim\epsilon\lesssim1$~GeV.
Here, we just show an example, not fixing the model parameters,
and have not calculated the secondary CR spectra like boron.
However, the future observations of short-lived CR radioactive
nuclei compositions can provide a clue to the nearby CR sources.
\par
%
%%%%%%%%%%%%%%%%%%%%
\begin{figure}
\plotone{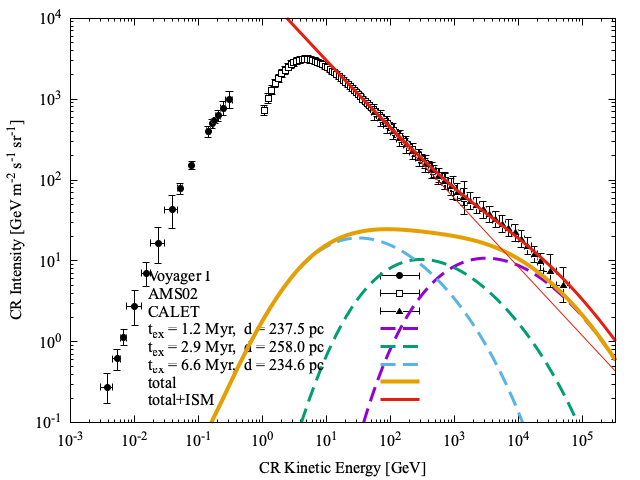}
\caption{The same as Figure~\ref{fig:GCR} but with different
source positions. The orange solid line represents the sum of
the contributions from three distinct sources as
$(t_{\rm age},d)=(1.2~{\rm Myr},~ 237.5~{\rm pc})$ (purple),
$(2.9~{\rm Myr},~ 258.0~{\rm pc})$ (green), and
$(6.6~{\rm Myr},~ 234.6~{\rm pc})$ (cyan).
The thick red line is the total contributions of the nearby sources
(solid orange) and distant sources expressed by a single power-law
component (thin red).}
\label{fig:GCR_far}
\end{figure}
%%%%%%%%%%%%%%%%%%%%
%
While the example in Figure~\ref{fig:GCR} would be extreme,
Figure~\ref{fig:GCR_far} shows a modest case with partially
contributing nearby sources.
In this case, the current position of the solar system
is outside the cavities of all three sources.
In this case, the dominant CR sources are outside the Local Bubble,
and the local sources contribute only to the softening and hardening
of the spectrum above 600 GeV.
Note that recent observations of external galaxies such as
NGC~628 reveal the galactic arms construction as a chain of
local bubbles \citep{watkins23}. Such a complicated structure
of the ISM would also be important to consider the origin of
Galactic CRs, especially for a larger diffusion coefficient.
We will extend our model to treat the CR compositions and
the more realistic ISM structures in the future.

\section{Summary}
\label{sec:summary}
We have studied the CR propagation around a CR source
with the CR hydrodynamical simulations, focusing on the dependence of
the CR diffusion coefficient.
When the diffusion coefficient
is suppressed compared to the standard value, the CR spectrum is
modified by the interaction with the ISM fluid.
CRs can expel the disk gas with
a total mass-loss rate of $\sim10~M_\odot~{\rm yr^{-1}}$
\citep{shimoda22a}.
This rate is comparable to the Galactic SFR,
consistent with the expected rate in the Galactic evolution scenarios
\citep{shimoda24}.
We have studied the effects
of CR heating and have found that the effects can be
tested by observations of atomic lines in the optical band
such as H$\alpha$ and [OIII]$\lambda5007$ \citep{fesen24}.
We have also demonstrated that a few nearby supernovae, which formed
the Local Bubble \citep{zucker22}, can be responsible for
the observed Galactic CR spectrum
around the Earth with a suppressed diffusion coefficient.
As a prompt conclusion, our model not only reproduces the observed local proton spectrum
by considering the interplay between escaping CRs and the ISM, but also
simultaneously provides a natural explanation for the H$\alpha$ and [OIII]
emissions in old SNRs. This dual success suggests that CR-induced heating
and dynamical effects play a more fundamental role in shaping both the ISM and CR environments
of our Galaxy than previously thought

%% Please use the acknowledgment and contribution environments. This will 
%% be anonomyized when the "anonymous" style option is used. 
\begin{acknowledgments}
%We thank all the people that have made this AASTeX what it is today.  This
%includes but not limited to Bob Hanisch, Chris Biemesderfer, Lee Brotzman,
%Pierre Landau, Arthur Ogawa, Maxim Markevitch, Alexey Vikhlinin and Amy
%Hendrickson. Also special thanks to David Hogg and Daniel Foreman-Mackey
%for the new {\tt\string modern} style design. Considerable help was provided via bug
%reports and hacks from numerous people including Patricio Cubillos, Alex
%Drlica-Wagner, Sean Lake, Michele Bannister, Peter Williams, Jonathan
%Gagne, Arthur Adams, Nicholas Wogan, Aaron Pearlman, Jeff Mangum, Mark Durre, Joel Ong, and Stephen Thorp.
The authors thank Y. Ohira for the fruitful discussions.
We are grateful the anonymous referee for comments that improved the paper.
This work is supported by the joint research program of the Institute for Cosmic Ray Research (ICRR), the University of Tokyo, and KAKENHI grant Nos. 22K03684, 23H04899, 24H00025, and 25K07352 (K.A.), 24K00677 (J.S.), and 25H00394 (S.I., and J.S.).

\end{acknowledgments}

\software{}

%% Appendix material should be preceded with a single \appendix command.
%% There should be a \section command for each appendix. Mark appendix
%% subsections with the same markup you use in the main body of the paper.
%%
%% Each Appendix (indicated with \section) will be lettered A, B, C, etc.
%% The equation counter will reset when it encounters the \appendix
%% command and will number appendix equations (A1), (A2), etc. The
%% Figure and Table counter will not reset.

\appendix

%% For this sample we use BibTeX plus aasjournalv7.bst to generate the
%% the bibliography. The sample7.bib file was populated from ADS. To
%% get the citations to show in the compiled file do the following:
%%
%% pdflatex sample7.tex
%% bibtext sample7
%% pdflatex sample7.tex
%% pdflatex sample7.tex

\bibliography{reference_paper}{}
\bibliographystyle{aasjournalv7}

%% This command is needed to show the entire author+affiliation list when
%% the collaboration and author truncation commands are used.  It has to
%% go at the end of the manuscript.
%\allauthors

%% Include this line if you are using the \added, \replaced, \deleted
%% commands to see a summary list of all changes at the end of the article.
%\listofchanges

\end{document}